\documentclass[12pt]{article}
\usepackage{epsf}
\usepackage[scanall]{psfrag}
\usepackage{graphicx}

\renewcommand \thesection{\arabic{section}.0}

\makeatletter
\newcounter {newsection}[section]
\renewcommand \thenewsection {\@arabic\c@section}
\renewcommand \thesubsection {\@arabic\c@newsection.\arabic{subsection}}
\newcounter {newnewsection}[section]
\renewcommand \thenewnewsection {\@Alph\c@section}
\makeatother

\def\bF{\mathbf{F}}
\def\bU{\mathbf{U}}
\def\bV{\mathbf{V}}
\def\bQ{\mathbf{Q}}
\def\bI{\mathbf{I}}

\renewcommand\appendix{\par
  \setcounter{section}{0}%
  \setcounter{subsection}{0}%
  \renewcommand\thesection{Appendix \Alph{section}}}

\begin{document}
\begin{flushleft}
Presented at the APS Centennial Meeting March 23, 1999, Atlanta,
Georgia; Session no.\ LB08. Available as an APS e-print
at http://publish.aps.org/eprint/. Paper number: aps1999feb12\underbar{~}001.
\end{flushleft}
\vskip1in

\begin{center}
{\Large Topological Constraints on Long-Distance}\\
{\Large Neutrino Mixtures}\\[3mm]
\hskip1in\vbox{\hbox{\large Gerald L. Fitzpatrick}
\hbox{\em PRI Research and Development Corp.}
\hbox{\em 12517 131 Ct. NE}
\hbox{\em Kirkland, WA 98034}
\hbox{pri@halcyon.com}}
\end{center}

\makeatletter
\renewcommand*\l@section[2]{%
  \ifnum \c@tocdepth >\z@
    \addpenalty\@secpenalty
    \addvspace{1.0em \@plus\p@}%
    \setlength\@tempdima{2.5em}%
    \begingroup
      \parindent \z@ \rightskip \@pnumwidth
      \parfillskip -\@pnumwidth
      \leavevmode \bfseries
      \advance\leftskip\@tempdima
      \hskip -\leftskip
      #1\nobreak\hfil \nobreak\hb@xt@\@pnumwidth{\hss #2}\par
    \endgroup
  \fi}
\makeatother

\begin{abstract}
\addcontentsline{toc}{section}{\numberline{}Abstract}
\noindent 
A new internal description of fundamental fermions (quarks and leptons),
based on a matrix-generalization ($\bF$) of the scalar fermion-number $f$,  
predicts that only 
three families of quarks and leptons, and their associated neutrinos ($\nu_e$, 
$\nu_\mu$ and $\nu_\tau$), exist.  Moreover, this description places important 
topological constraints on neutrino mixing. For example, with respect to $\bF$, 
the topology of the
$\nu_e$ ($\nu_\mu$ or $\nu_\tau$) is that of a cylinder (M\"obius strip). 
Assuming that a change 
in topology during neutrino-neutrino transitions is suppressed (e.g., one 
cannot continuously deform a donut into a sphere), while neutrino-neutrino
transitions without 
topology-change are (relatively) enhanced, one may have an explanation for 
recent short-distance experimental observations of (nearly) maximal 
$\nu_\mu$-$\nu_\tau$
mixing at the Super  Kamiokande.  To test this idea, I was able to 
use simple topological arguments to deduce a matrix describing long-distance 
neutrino mixtures, which is \emph{identical} to that proposed by Georgi and 
Glashow on different grounds. Experimental confirmation of this prediction
would strongly support the new description of fundamental fermions, which
requires, among other things, that
the $\nu_e$ and 
($\nu_\mu$  or $\nu_\tau$) neutrinos
start life as topologically-distinct quantum objects.
\end{abstract}
\newpage

\section{Introduction}

Except where explicitly prevented by some ``absolute'' conservation law
(e.g., the conservation of electric charge or spin angular momentum),
quantum mechanics generally permits transitions between states having
\emph{different} topologies [1]. While a change in topology may be energetically
(or otherwise) inhibited, unavoidable quantum fluctuations are expected to
catalyze such processes. Hence, there is always the possibility of
\emph{mixing} between otherwise similar states having \emph{distinct}
topologies [2].

Recently, a new internal description of fundamental fermions (quarks and 
leptons) was proposed [3].  The new description is based on a 
matrix-generalization $\bF$ of the scalar fermion number $f$.
One of the main predictions of the new description is that only three 
families of quarks and leptons exist---hence that there are only three low-mass 
neutrino flavors, namely, the $\nu_e$, $\nu_\mu$  and $\nu_\tau$ neutrinos.  
Moreover, because of 
the way fundamental fermions are represented by certain geometric objects in 
the space on which the matrix transformation $\bF$ acts [Ref.\ 3, p.\ 57 and
pp. 85--87, Ref.\ 4, 5 and Ref.\ 6, pp.\ 244--255],
the $\nu_e$ and ($\nu_\mu$  or $\nu_\tau$)
neutrinos are found to have \emph{different} topologies with respect to $\bF$. 
This fact turns out to have important implications for neutrino mixing.

With respect to the matrix transformation F,  the topology of both the 
$\nu_\mu$  and $\nu_\tau$
is found to be that of a M\"obius strip [Ref.\ 6, p.\ 143].  By contrast,  
the topology of the $\nu_e$
(with respect F) is that of a cylinder.
And because a change in topology 
during transitions tends to be suppressed (e.g., one cannot continuously 
deform a donut into a sphere) the foregoing topological distinctions between 
neutrinos may help explain recent experimental observations of (nearly) 
maximal $\nu_\mu$-$\nu_\tau$  
mixing at the Super Kamiokande facility [7]. The purpose of 
this paper is to determine the constraints that these topological distinctions 
place on long-distance neutrino mixtures, and by implication, short-distance 
neutrino mixtures as well.

\section{Topological Constraints on Neutrino Mixtures}
In all conventional weak transitions, including neutral-current weak
transitions, the $e$ $\mu$ and $\tau$ quantum numbers are additively
conserved. Although the observed
neutrino mixing [7] violates these conservation laws,  all
transitions of neutrinos from one flavor to another (i.e.,
$\nu_l\to\nu_{l'}$,
$l\ne l'$) will tend to be suppressed.
In this paper we implicitly assume that this degree, and kind,
of suppression is the \emph{same} for all $\nu_l\to \nu_{l'}$, and
therefore, that this effect can be completely \emph{ignored} when discussing
(relative) topological constraints placed on long-distance neutrino
mixtures.

\setcounter{newsection}{2}

\subsection{Additional constraints on the matrix $M$}
Given that the $\nu_e$ ($\nu_\mu$ or $\nu_\tau$) neutrino 
has the topology of a cylinder
(M\"obius strip) with respect to the internal transformation $\bF$,
and assuming that topological constraints are the
\emph{primary} determinants of the matrix $M$ describing long-distance neutrino 
mixtures, the \emph{form} of $M$ is easily determined [8].
Moreover, these same
topological constraints can be used to place numerical bounds on the
components of $M$. To accomplish these results we need only apply the
following very general principle to neutrino-neutrino transitions:
\medskip

\emph{All other things being equal, any neutrino flavor $\nu_l$, {\rm(}i.e.,
$\nu_e$, $\nu_\mu$ or $\nu_\tau${\rm)} that undergoes neutrino-neutrino
transitions involving a change in topology, will tend to be suppressed,
while neutrino-neutrino transitions involving no change in topology will
tend to be {\rm(}relatively{\rm)} enhanced.}
\medskip

To this principle we add the following corollary,
\medskip

\emph{All other things being equal, because the $\nu_\mu$ and $\nu_\tau$ have
the same topology, they will act the same way in all neutrino-neutrino
transitions.}
\medskip

Given these principles, we immediately have the following two constraints on
long-distance neutrino mixtures:
\medskip

A. No matter what neutrino flavor $(\nu_l)$ and topology one starts with at
some distant source (say a supernova), by the time the neutrino 
mixture reaches its
``equilibrium'' state, it should contain \emph{equal} fractions of $\nu_\mu$
and $\nu_\tau$, because these neutrinos have the \emph{same} topology.
\medskip

B. Because the $\nu_\mu$ and $\nu_\tau$ have the \emph{same} topology, if one
starts out with \emph{either} a pure $\nu_\mu$ \emph{or} a pure $\nu_\tau$
source, one should end up with the \emph{same} long-distance equilibrium
mixture of $\nu_e$, $\nu_\mu$ and $\nu_\tau$.
\medskip

To describe A) and B) in conventional terms [9], let $D_l$ be the number of
detected neutrinos of type $\nu_l$ and $B_l$ be their number at ``birth''
at some distant source. Then
\begin{equation}
\{D_e, D_\mu, D_\tau\}=M\{B_e, B_\mu, B_\tau\},
\end{equation}
where $M$ is a $3\times 3$ matrix of fractional coefficients describing the
long-distance neutrino
mixture, and $\{\;,\;,\;\}$ are column ``vectors''. By definition, all rows
and columns of $M$ must sum to unity (100\%), which means that $D_l$ and
$B_l$ must satisfy the constraint
\begin{equation}
D_e+D_\mu+D_\tau=B_e+B_\mu+B_\tau.
\end{equation}

\subsection{The form of the matrix $M$}

Constraints A) and B), together with Eqs.\ (1) and (2), dictate that the
matrix $M$ describing long-distance neutrino mixtures must have the
symmetrical form
\begin{equation}
\left(\begin{array}{c}
D_e\\
D_\mu\\
D_\tau\end{array}\right) = \left(
\begin{array}{ccc}
a&b&b\\
b&c&c\\
b&c&c\end{array}\right) 
\left(\begin{array}{c}
B_e\\
B_\mu\\
B_\tau\end{array}\right).
\end{equation}
While topological constraints alone cannot determine the exact numerical
values of the matrix elements $a$, $b$ and $c$, they can place numerical
bounds on these quantities. For example, because the $\nu_e$ is
\emph{inhibited} by its topology from turning into a $\nu_\mu$ or
$\nu_\tau$, we naturally expect that (See Fig.\ 2.1)
\begin{equation}
a\ge b.
\end{equation}
Similarly, because the $\nu_\mu$ and/or the $\nu_\tau$ are \emph{inhibited} by
their topology from turning into a $\nu_e$, we naturally expect that (See
Fig.\ 2.1)
\begin{equation}
c\ge b.
\end{equation}

Equations (4) and (5), together with the requirement that
all rows and columns of the matrix $M$ in (1) and (3) sum to unity (i.e.,
$a+2b=b+2c=1$), further leads to the following bounds 
on
$a$, $b$ and $c$:
\begin{equation}
(\frac{1}{3}\le a\le 1),\qquad (0\le b\le \frac{1}{3}),\qquad (\frac{1}{3}\le
c\le \frac{1}{2}).
\end{equation}
Moreover, since $a+2b=b+2c$, the arithmetic mean of $a$ and $b$ is $c$,
i.e., $(a+b)/2=c$, which means that $c$ lies half-way 
between $a$ and $b$. Hence,
the matrix elements $a$, $b$ and $c$ are subject to the constraint
\begin{equation}
a\ge c\ge b.
\end{equation}
Equations (1) and (3) determine the general \emph{form} of $M$, while (6)
places bounds on its components, whose relative \emph{magnitudes} are further
constrained by (7).

Finally, because rows and columns of $M$ sum to unity, $M$ may be expressed
in terms of the \emph{single} parameter $a$ as
\begin{equation}
M(a)=\frac{1}{4} \left( \begin{array}{lll}
4a, & 2(1-a), & 2(1-a) \\
2(1-a), & (1+a), & (1+a) \\
2(1-a), & (1+a), & (1+a) \end{array}\right).
\end{equation}

\begin{figure}
\begin{center}
\begin{tabular}{ll}
Transitions \emph{without} & Transitions \emph{with} \\
topology change & topology change \\
(5 matrix elements) & (4 matrix elements)\\
\begin{psfrags}
\includegraphics[width=2.5in]{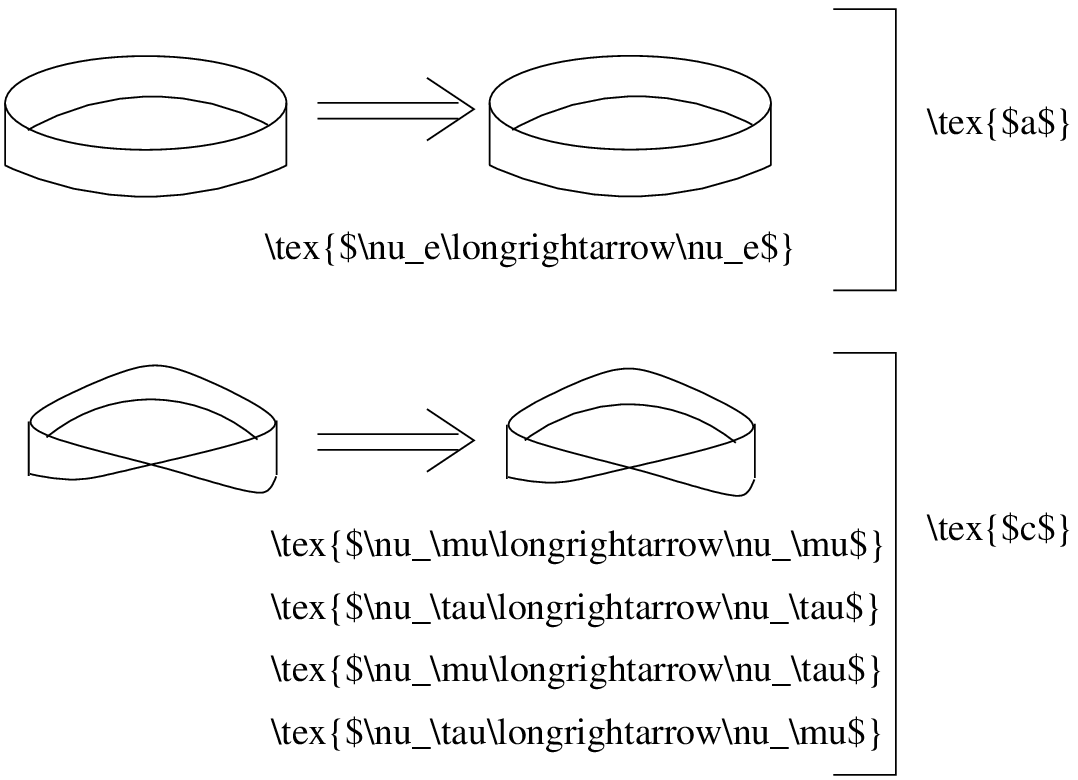} 
\end{psfrags}
& 
\begin{psfrags}
\includegraphics[width=2.5in]{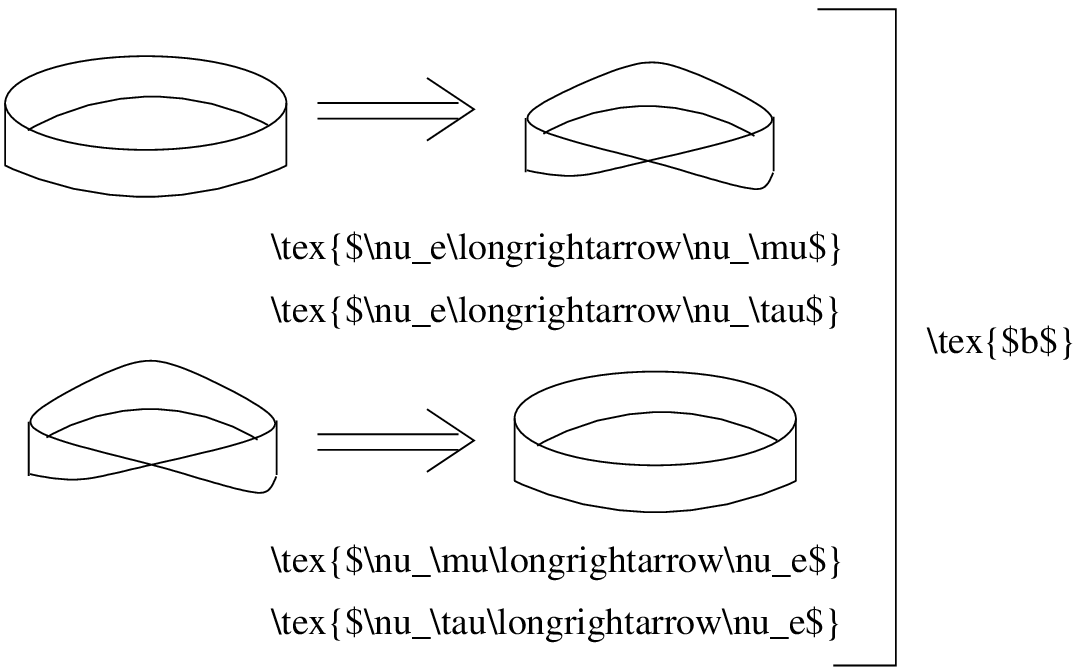}
\end{psfrags}
\end{tabular}
\end{center}
\noindent{\bf Figure 2.1. Matrix elements $a$, $b$ and $c$ describing
long-distance neutrino mixtures [see Eq.\ (3)].} Neutrino-neutrino
transitions without topology change are ``preferred'' relative to
neutrino-neutrino transitions involving a change in topology. That is, the
matrix elements $a$ and $c$ are both greater than $b$. The symbol
\lower3pt\hbox{\begin{psfrags}\includegraphics[width=.3in]{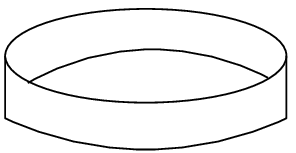}
\end{psfrags}}
(~\lower2pt\hbox{\begin{psfrags}\includegraphics[width=.3in]{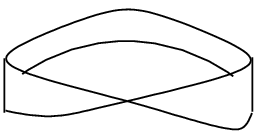}
\end{psfrags}}$\,$)
for a cylinder
(M\"obius strip) is used to represent the topology of the $\nu_e$ ($\nu_\mu$
or $\nu_\tau$) with respect to the internal transformation $\bF$. Note that
the matrix elements $b$ and $c$ are unaffected by an exchange of the
subscripts $\mu$ and $\tau$.
\end{figure}

\subsection{The Georgi-Glashow matrix $M(\frac{1}{2})$}

The \emph{form} of $M$ specified in (3) and (8), together with the constraints
provided by (6) and (7), are completely consistent with the matrix proposed
by Georgi and Glashow [9] to describe long-distance neutrino mixtures, namely,
\begin{equation}
M({\textstyle\frac{1}{2}}) = \frac{1}{8}\left(\begin{array}{ccc}
4 & 2 & 2 \\
2 & 3 & 3 \\
2 & 3 & 3 \end{array}\right).
\end{equation}
This matrix can be viewed in two different, but apparently complementary,
ways as being
\begin{enumerate}
\item[1)]The direct result of an assumed form for the neutrino mass-matrix
(and associated mixing parameters), and other factors dictated by requiring
consistency with the standard-model description of neutrinos, including all
known experimental properties of neutrinos. This is the point of view
espoused in [9],
\end{enumerate}
and/or
\begin{enumerate}
\item[2)]The result of a balance between opposing topological and quantum
``forces'' acting on neutrinos. This is the point of view espoused in the
present paper.
\end{enumerate}

If 1) and 2) above are to be compatible points of view, then it follows that
such things as the neutrino mass-matrix is, in some sense, the result of an
interplay between opposing topological and quantum ``forces'' acting on
neutrinos. This conclusion would gain support if it were possible to
\emph{derive} the matrix $M$ without reference to the approach taken in
[9]. Such a derivation is presented in Section 3.0.  
But, before turning to the derivation of $M$, we must first examine the
boundary conditions on this matrix.

\subsection{Boundary conditions on $M(a)$}

Consider the general matrix $M(a)$ defined on the \emph{boundaries} of the
physically accessible region $(\frac{1}{3}\le a\le 1)$. When
$a=a_{\frac{1}{3}}=\frac{1}{3}$ (See Eq.\ 8 and Ref.\ 10)
\begin{equation}
M({\textstyle\frac{1}{3}})=\frac{1}{3} \left(\begin{array}{ccc}
1 & 1 & 1 \\
1 & 1 & 1 \\
1 & 1 & 1 \end{array}\right).
\end{equation}
In this case, topological constraints are effectively nonexistent---topology 
change in neutrino-neutrino transitions is completely \emph{unrestricted}
$(a=b=c=\frac{1}{3})$. That is, quantum fluctuations, which act to catalyze
topology-change, are completely dominant when $a=\frac{1}{3}$. Figure 2.2
illustrates the situation.
\def\a{$\left(\begin{array}{l}
        \hbox{``Forces'' acting to oppose}\\
        \hbox{topology-change}\end{array}\right)$}
\def\b{(No topology change)}
\def\c{(Unrestricted topology change)}
\def\d{$\left(\begin{array}{l}
\hbox{``Forces'' catalyzing topology-change, i.e.,}\\
\hbox{quantum fluctuations}\end{array}\right)$}
\begin{figure}[!ht]
\begin{center}
\begin{psfrags}
\includegraphics[width=6in]{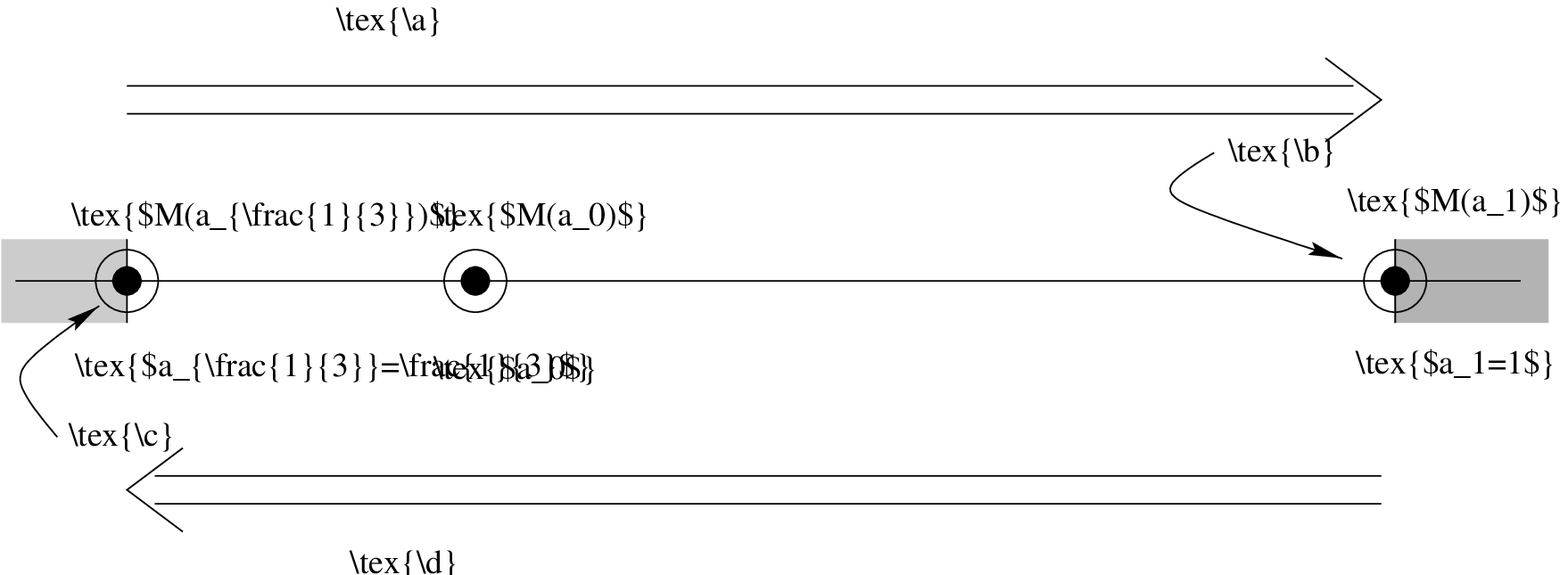}
\end{psfrags}
\end{center}
\noindent{\bf Figure 2.2. The balance between opposing quantum and
topological ``forces''.} The parameter $a$ varies from $a_{\frac{1}{3}}=
\frac{1}{3}$ (no effective topological constraints on neutrino mixing) 
to $a_1=1$
(maximal topological-constraints in neutrino mixing). The shaded regions are
physically inaccessible. Owing to unavoidable
quantum-fluctuations, the ``forces'' acting to oppose topology-change are,
themselves opposed, by quantum ``forces'' acting to catalyze topology-%
change. These two opposing ``forces'' must seek a stable ``equilibrium'' at
some point $a=a_0$ located between $a_{\frac{1}{3}}$ and $a_1$. Thus the
matrix $M(a_0)$ represents a ``balance'' between the extreme conditions
$M(a_{\frac{1}{3}})$ and $M(a_1)$.
\end{figure}

When $a=a_1=1$ (See Eq.\ 8 and Ref.\ 10)
\begin{equation}
M(1)=\frac{1}{2} \left(\begin{array}{ccc}
2 & 0 & 0 \\
0 & 1 & 1 \\
0 & 1 & 1 \end{array}\right).
\end{equation}
In this case, topological constraints are maximal---topology-change in
neutrino-neutrino transitions is completely \emph{prevented} $(a=1, b=0,
c=\frac{1}{2})$. That is, quantum fluctuations, which would otherwise
catalyze topology-change, are effectively overwhelmed by topological
energy (or other topological) ``barriers'' when $a=1$. Figure 2.2
illustrates the situation.

Now, because of (8, 10 and 11), any matrix $M(a)$ defined on the interval
$(\frac{1}{3}\le a\le 1)$ can be written as a function of the single
parameter $a$ as 
\begin{equation}
M(a) = \frac{3}{2}(1-a) M({\textstyle\frac{1}{3}})+\frac{3}{2}(a-\frac{1}{3})M(1)
\end{equation}
where
\begin{equation}
\frac{3}{2}(1-a) + \frac{3}{2}(a-\frac{1}{3})=1.
\end{equation}
Clearly, the fractions $\frac{3}{2}(1-a)$ and $\frac{3}{2}(a-\frac{1}{3})$,
determine the fraction of $M(a)$, which depends on the topology-changing
piece $M(\frac{1}{3})$ and the topology-main-\break
taining piece $M(1)$,
respectively.

In the case of the Georgi-Glashow matrix in (9) we find
\begin{equation}
M({\textstyle\frac{1}{2}}) = \frac{3}{4}M({\textstyle\frac{1}{3}})+
\frac{1}{4}M(1).
\end{equation}
Presumably, this particular three-to-one ``mixture'' of 
$M(\frac{1}{3})$ and $M(1)$, respectively,
represents the equilibrium condition between opposing topology-changing
and topology-maintaining ``forces'', respectively.

\section{A Simple Calculation of $a$, $b$ and $c$.}
According to (3) and (8) the matrix elements $a$, $b$ and $c$ are given by
\begin{equation}
\left.\begin{array}{rcl}
a & = & a \\[3mm]
b & = & \frac{1}{2}(1-a) \\[3mm]
c & = & \frac{1}{4}(1+a)\end{array}\right\}\lower.2in\hbox{.}
\end{equation}
Focus attention on the pair of matrix elements $a$ and $b$, which \emph{bracket} $c$.
 Now,
topology-changing ``forces'' (quantum fluctuations) constantly ``attempt'' to
\emph{maximize} $b$ at the expense of $a$ (and $c$), while at the same time,
topology-maintaining ``forces'' (energy barriers ?) constantly ``attempt'' to
maximize $a$ (and $c$) at the expense of $b$ (See Figures 2.1 and 2.2). 

Since $a$ and $b$ are \emph{both} functions of $a$, a ``compromise'' between
these opposing ``forces'' should be reached when the product $f(a)=ab$ is
\emph{maximized} [11] with respect to variations of $a$. From (15)
one has
\begin{equation}
f(a)=\frac{1}{2}a(1-a),
\end{equation}
which possesses a \emph{maximum} at a value of $a$ equal to the
\emph{harmonic mean} [12] of the boundary values $a_{\frac{1}{3}}$ and $a_1$ 
(See Fig.\ 2.2), namely,
\begin{equation}
a_0 = \frac{2a_{\frac{1}{3}}\cdot a_1}{a_{\frac{1}{3}}+a_1} = \frac{1}{2}.
\end{equation}
And, \emph{this value of 
{\rm a} 
is precisely the value required to yield the
Georgi-Glashow matrix} $M(\frac{1}{2})$ of (9). However, one other 
conceivable
solution for $a_0$ exists.

The function $g(a)=abc$ also possesses a \emph{maximum}, which occurs at a
value of $a$ equal to the \emph{geometric mean} of the boundary values
$a_{\frac{1}{3}}$ and $a_1$ (See Fig.\ 2), namely,
\begin{equation}
a_0=\sqrt{a_{\frac{1}{3}}\cdot a_1} = \frac{\sqrt{3}}{3}=0.577.
\end{equation}
Equations (17) and (18) represent the \emph{only} physically acceptable
solutions involving extremal functions that depend on simple products of the
matrix elements $a$, $b$ and $c$ (i.e., $ab$, $bc$, $ac$ and $abc$).

For example, while the function $h(a)=b\cdot c$ possesses a \emph{maximum},
this occurs at a value of $a$ equal to zero, which point lies \emph{outside}
the physically acceptable region $(a_{\frac{1}{3}}\le a\le a_1)$. And, the
function $i(a)=a\cdot c$ possesses a \emph{minimum} instead of a maximum.
Moreover, this occurs at an ``unphysical'' value of $a$, namely,
$a_0=-\frac{1}{2}$.

In summary, without further constraints, (17) and (18) would seem to be
equally likely solutions. Is there some way to decide which solution nature
\emph{should} choose?

\setcounter{newsection}{3}

\subsection{Discussion}
Three reasons can be cited in favor of (17) and $M(\frac{1}{2})$. These are
\begin{enumerate}
\item From the standpoint of \emph{simplicity} and \emph{parsimony} the
\emph{quadratic} $f(a)=ab$ is obviously simpler and more parsimonious than
the \emph{cubic} $g(a)=a\cdot b\cdot c$. For example, 
the quadratic is symmetrical 
about its maximum, while the cubic is asymmetrical about its maximum. 
The function $f(a)$ is also simpler and
more parsimonious than $g(a)$, in the sense that $g(a)$ has \emph{two}
factors $a$ and $c$ 
that describe the effect of topology-maintaining ``forces'', while
$f(a)$ has only \emph{one} such factor $a$. 
Hence, for reasons of simplicity
and parsimony, (17) and $M(\frac{1}{2})$ are favored over (18) and
$M(0.577)$, respectively.
\item From [9] the matrix $M(\frac{1}{2})$ is known to be consistent with a
reasonable form for the neutrino mass-matrix and associated mixing-%
parameters. The choice of (18) and $M(0.577)$ would require changes in these
quantities that may, or may not, be consistent with the presently known
experimental
properties of neutrinos. Hence, (17) and $M(\frac{1}{2})$ appear to be
favored for this reason over (18) and $M(0.577)$, respectively.
\item In Appendix A we present a crude ``derivation'' of the matrix $M$
based on a
harmonic-oscillator description of the balance between topological
and quantum ``forces.'' This description also seems to favor (17) and
$M(\frac{1}{2})$ over (18) and $M(0.577)$, respectively.
\end{enumerate}

It is remarkable that, in the absence of an in-depth understanding of
topology-maintaining and topology-changing ``forces'' (See Fig.\ 2.2), 
it has nevertheless
proven possible to obtain an \emph{exact} calculation (see qualifying
remarks above) of the matrix elements $a$, $b$ and $c$, which describe
long-distance neutrino mixtures. This apparent success strongly supports the
main premise of this paper, which is that topological constraints play a
\emph{major} role in describing both short- and long-distance neutrino
mixtures.

\section{Conclusions}

Topological constraints appear to play a major role in determining the 
nature of both short- and long-distance neutrino mixtures.  Given that the
$\nu_e$ and ($\nu_\mu$ or $\nu_\tau$) neutrinos have distinct topologies,
and assuming that 
topology changes are suppressed in neutrino-neutrino 
transitions, while neutrino-neutrino 
transitions without topology-change are relatively enhanced--one easily 
determines both the form of the matrix describing
long-distance neutrino mixtures, and even
fixes the numerical magnitudes of the individual matrix elements.
If the predicted matrix (9),
or one close to it [i.e., $M(0.577)$], is eventually confirmed by observations 
of neutrinos from distant astronomical sources,
this will strongly
support the new description of fundamental fermions, which 
requires, among other things, that
the $\nu_e$ and ($\nu_\mu$  or $\nu_\tau$) neutrinos start life
as topologically-distinct quantum objects [3--5].

\section{Appendix A. A Toy Model for Calculating $M(a_0)$.}

\setcounter{section}{1}

\renewcommand{\theequation}{\thenewnewsection\arabic{equation}}
\setcounter{equation}{0}

Imagine that the parameter $a$ describes a dynamical ``system'' associated
with neutrino-neutrino mixtures. In particular, this system has a
\emph{single} degree of freedom (is constrained to move on a straight line),
which is further bounded by $a_1$ and $a_{\frac{1}{3}}$ $(a_{\frac{1}{3}}\le
a\le a_1)$. Treating $a$ as a kind of ``generalized coordinate'' we are led
to consider the ``generalized forces'' acting on the ``system'' and to
determine its \emph{equilibrium} ``position'' $a_0$ (See Fig.\ 2.2). Quantum
fluctuations, in effect, ``force'' the ``system,'' as described by $a$ from
the value $a=a_1$ toward $a=a_{\frac{1}{3}}$. At the same time, topological
energy (or other topological) ``barriers'' effectively ``force'' the system as described
by $a$, from the value $a=a_{\frac{1}{3}}$ toward $a=a_1$. These two
opposing ``forces'' must reach a compromise, i.e., \emph{stable
equilibrium}, at some value $a$ (call it $a_0$), which lies somewhere
between $a_{\frac{1}{3}}$ and $a_1$. This value $(a_0)$ we take to describe
long-distance neutrino mixtures via the matrix $M(a_0)$. 

To calculate $a_0$, let us model the ``system'' using a simple (classical)
harmonic-oscillator ``potential'' (See Ref.\ 13 and Figure A.1),
\begin{equation}
V(x)=\frac{k}{2}x^2,
\end{equation}
where the Hooke's-law ``restoring force'' is
\begin{equation}
F=-kx=\frac{-\partial V(x)}{\partial x},
\end{equation}
and $x=(a-a_0)$ is the ``displacement'' of the system from its (stable)
equilibrium ``position'' $a=a_0$. 

\def\a{$F(a_1)=-\frac{\partial V(a)}{\partial a}\biggm|_{a_1}
=-2(a_1-a_0)$}
\def\b{$\begin{array}{rcl}
F(a_{\frac{1}{3}}) & = & -\frac{\partial V(a)}{\partial a}
\biggm|_{a_{\frac{1}{3}}}\\
& = & -2(a_{\frac{1}{3}}-a_0)\end{array}$}
\def\c{$V(a_{\frac{1}{3}})=\left(\frac{a_{\frac{1}{3}}}{a_1}\right)^2V(a_1)$}

\begin{figure}[!ht]
\begin{center}
\begin{psfrags}
\includegraphics[width=6in]{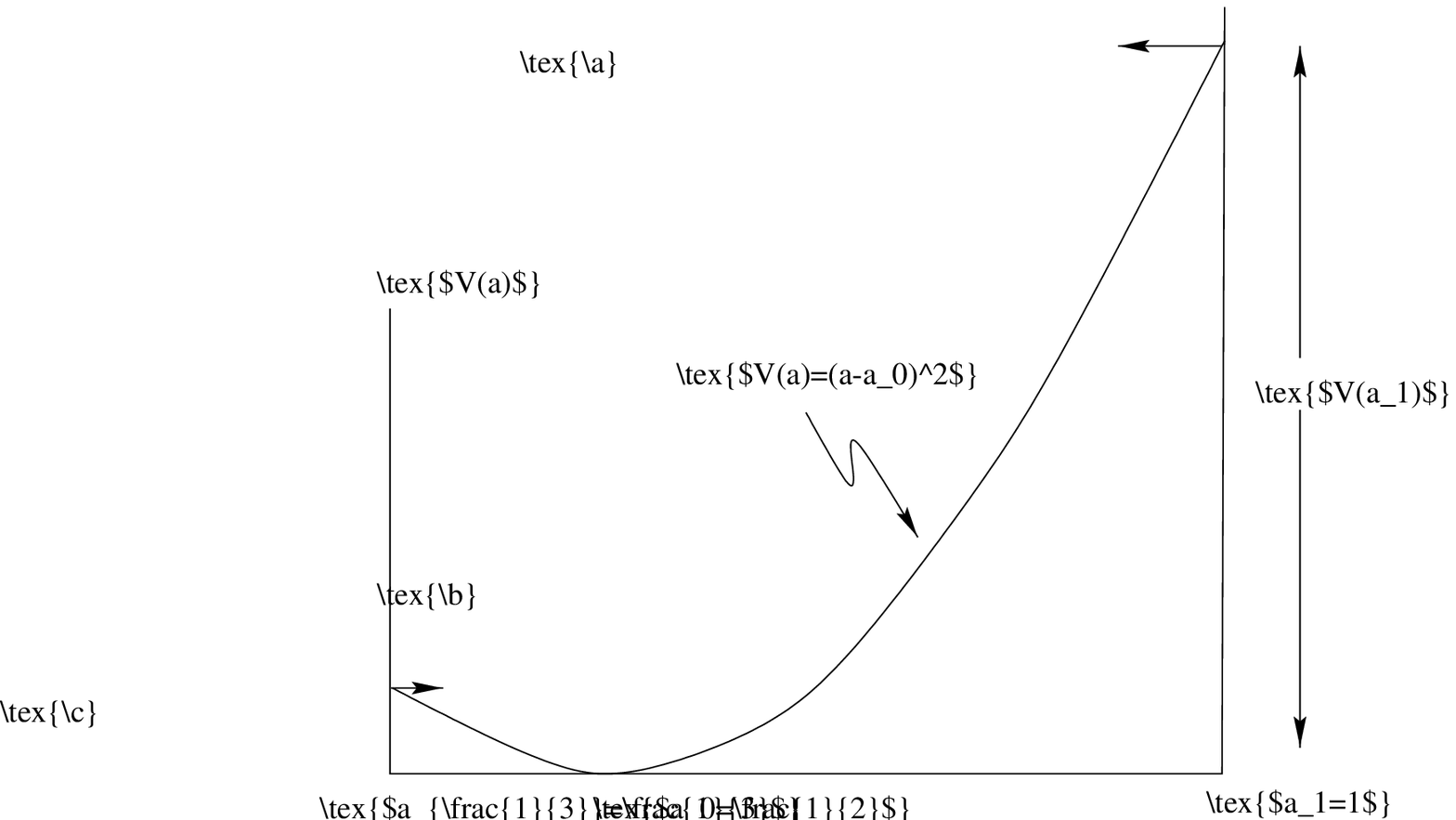}
\end{psfrags}
\end{center}
\noindent{\bf Figure A1. The ``potential'' $V(a)$ describing the opposing
quantum and topological ``forces''.} These ``forces'' balance one another at
the (stable) equilibrium position $a_0$. Here, $k=2$, $V(x)=\frac{k}{2}x^2 =
(a-a_0)^2$, and the ``restoring force'' is $F(a)=-2(a-a_0)$.
\end{figure}

From (A1) we have
\begin{equation}
\frac{\partial V(x)}{\partial x}=k(a-a_0).
\end{equation}
Clearly, if $k(a-a_0)$ were known on each boundary, then $k$ and $a_0$ could
be calculated. To do this, note first that $\partial V(x)/\partial 
x$ must be a number with
the ``units'' of $a$ (assuming $k$ is dimensionless). And, since the only
number available to describe the system on the boundary of $a$, is $a$
itself, we make the following assumptions
\begin{eqnarray}
\frac{\partial V(x)}{\partial x}\biggm|_{a_1} & = & k(a_1-a_0)=+a_1;
\;\;\;\;\;\;\, a_1>a_0\\
\frac{\partial V(x)}{\partial x}\biggm|_{a_{\frac{1}{3}}} & = & k(a_{\frac{1}{3}}-a_0) =
-a_{\frac{1}{3}};\;\;\;\; a_{\frac{1}{3}}<a_0.
\end{eqnarray}
From (A4) and (A5) we can easily deduce the unknowns $k$ and $a_0$. One
finds 
\begin{equation} 
k=\frac{(a_1+a_{\frac{1}{3}})}{(a_1-a_{\frac{1}{3}})} = 2,
\end{equation}
and 
\begin{equation}
a_0= \frac{2(a_{\frac{1}{3}}\cdot a_1)}{(a_{\frac{1}{3}}+a_1)} = \frac{1}{2}\,.
\end{equation}
Both $k$ and $a_0$ depend only on the values of $a$ on the
boundaries. In particular,  $a_0$ is found to be the
\emph{harmonic mean} of $a_{\frac{1}{3}}=\frac{1}{3}$ and $a_1=1$.

\setcounter{section}{5}

\section{References and Footnotes}

\noindent
[1]  A.\ P.\ Balachandran, ``Bringing Up a Quantum Baby,'' 
quant-ph/9702055.
\medskip

\noindent
[2] G.\ Holzwarth, ``Formation of Extended Topological-Defects During 
Symmetry-Breaking Phase Transitions in $O(2)$ and $O(3)$ Models'',
hep-ph/9901296. Analogous examples of fluctuation-induced topology-change in
macroscopic objects (e.g., destruction of topological objects due to thermal
fluctuations at phase transitions) abounds. For example, otherwise
persistent (``conserved'') \emph{topological defects} in crystals can be
destroyed by raising the temperature sufficiently (i.e., by melting the
crystal). Similarly, otherwise persistent (``conserved'') magnetic
\emph{flux-tubes} in Type II superconductors and/or \emph{vortices} in a
superfluid, can both be destroyed by raising the temperature above the
critical temperature $T_c$. And, conversely, topological defects are always
created when such macroscopic systems first condense (or crystallize) as the
temperature is lowered. We imagine that something roughly similar can happen
when quantum-fluctuations (vacuum fluctuations) act on otherwise very
similar quantum states (i.e., same charge, spin and 
nearly identical mass) that also happen
to start life as distinct topological-objects. That is, we are assuming
that, if not prevented by some conservation law, transitions between such
states (e.g., $\nu_e$ and $\nu_\mu$ neutrinos) 
will be \emph{catalyzed} by quantum fluctuations.
\medskip

\noindent
[3]  Gerald L. Fitzpatrick, \emph{The Family Problem-New Internal Algebraic
and
Geometric Regularities}, Nova Scientific Press, Issaquah, Washington, 1997.
Additional information: {\tt http://physicsweb.org/TIPTOP/} or\hfill\break
{\tt http://www.amazon.com/exec/obidos/ISBN=0965569500}.
In spite of the many successes of the standard model of particle physics,
the
observed proliferation of matter-fields, in the form of ``replicated"
generations or families, is a major unsolved problem.  In this book
a new organizing principle for fundamental fermions is proposed, i.e.,
a minimalistic
``extension" of the standard model based, in part, on the Cayley-Hamilton
theorem for matrices.   To introduce (internal) global degrees
of freedom that are capable of distinguishing all observed flavors, the
Cayley-Hamilton theorem is used
to generalize the familiar standard-model concept of
scalar fermion-numbers $f$
 (i.e., $f_m=+1$ for all fermions and $f_a=-1$ for
all
antifermions).  This theorem states that every (square) matrix satisfies its
characteristic equation. Hence, if $f_m$ and $f_a$
are taken to be the eigenvalues
of some real matrix $\bF$ (a ``generalized fermion  number"), it follows
from
this theorem that both $f$ and $\bF$ are square-roots of unity.  Assuming
further that the components of both $\bF$ and its eigenvectors are global
charge-like quantum observables, and that $\bF$ ``acts" on a (real) vector
2-space, both the form of $\bF$
and the 2-space metric are determined.  One finds that the 2-space has a
non-Euclidean or ``Lorentzian" metric, and that various associated 2-scalars
serve as global flavor-defining ``charges," which can be identified with
charges such as strangeness, charm, baryon and lepton numbers etc..  Hence,
these global charges can be used to describe individual flavors (i.e.,
flavor
eigenstates), flavor doublets and families.  Moreover, because of the
aforementioned non-Euclidean constraints and certain standard-model
constraints, one finds
that these global charges are effectively- ``quantized" in
such a way that families are replicated.  Finally, because these same
constraints dictate that there are only a limited number of values these
charges can assume, one finds that families, and their associated neutrinos,
always come in ``threes."
\medskip

\noindent
[4]  The eigenvectors $\bQ$ of $\bF$ (i.e., $\bF\bQ=f\bQ$ where $f$ is the
scalar fermion-number), together with certain pairs of linearly independent
vectors ($\bU$ and $\bV$) that resolve $\bQ$ (i.e., $\bQ=\bU+\bV$), namely,
various non-Euclidean vector ``triads'' $(\bQ, \bU, \bV)$---these are the
analogs of Euclidean triangles---serve to represent flavor-doublets in terms
of a pair of quark or lepton flavor-eigenstates as follows:
\[
|\hbox{``up''}\rangle = |q_1, u_1, v_1, {\mathbf{Q}}^2, {\mathbf{U}}^2,
2\bU\cdot\bV\rangle
\]
and
\[
|\hbox{``down''}\rangle = |q_2, u_2, v_2, {\mathbf{Q}}^2,
{\mathbf{U}}^2, 2\bU\cdot\bV\rangle.
\]
Here, $\bQ=\{q_1, q_2\}$, $\bU=\{u_1, u_2\}$ and $\bV=\{v_1,v_2\}$ are column-%
vectors and their components $q_1, q_2, u_1, u_2$, $v_1$ and $v_2$,
together with the non-Euclidean scalar products $\bQ^2$, $\bU^2$ and
$\bV^2$, are various global mutually-commuting flavor-defining charge-like
quantum numbers.
When we refer to the ``topology'' of a particular neutrino flavor-eigenstate
(e.g., the $\nu_e$) we are referring to the topology of the corresponding
\emph{vector triad} $(\bQ, \bU, \bV)$, with respect to the internal
transformation $\bF$. And, because $\bF$ generates the M\"obius group $Z_2$
(i.e., $\bF^2=\bI_2)$, those vector ``triads'' that are left unchanged by
$\bF$, have the topology of a \emph{cylinder}, whereas vector triads that are
changed by $\bF$ (but obviously not changed by $\bF^2=\bI_2$), have the
topology of a \emph{M\"obius strip}. And, as it turns out, the neutrino
flavor $\nu_e$ ($\nu_\mu$ or $\nu_\tau$) corresponds to a vector triad 
having the topology of a cylinder (M\"obius strip) with respect to $\bF$.
\medskip

\noindent
[5] H.\ Umezawa, \emph{Advanced Field Theory-Micro, Macro and Thermal
Physics}, Am.\ Inst.\ of Physics, New York, 1993, pp.\ 128--129. An
\emph{inexact}, but nevertheless very suggestive, analogy can be drawn
between our 2D (non-Euclidean) charge-vectors $\bQ=\{q_1, q_2\}$,
and the 2D (Euclidean) topological-``charge'' vector called a Burgers
vector. The \emph{quantized} Burgers vector of a dislocation or defect
in a 2D crystal lattice, is defined as the net
number of extra rows and columns one encounters while traversing a closed
path around the defect, expressed as a vector $\{$columns, rows$\}$.  In
general, defects are discontinuities or ``tears'' in some order-%
parameter field. In this particular case, the defect is a topological
line-singularity in the crystalline order. 
Because the topological ``charge'' (Burgers vector)
is \emph{conserved} as the associated defect migrates through the
crystal lattice, the associated defect is also conserved within the lattice. 
Similarly, our \emph{quantized} charge-vector $\bQ$ is conserved
$(\Delta\bQ=0)$, as the associated fundamental-fermion (quark or
lepton) moves through spacetime. Conservation of the vector $\bQ$ leads to
the identification of $\bQ^2$ with the ``conserved'' baryon- or lepton-number.
Hence, we can think of the vector $\bQ$ as being a topological ``charge'',
which describes certain aspects of a
topological ``defect'' (fundamental fermion) in some spacetime
order-parameter field. The geometric
object we call a vector triad, namely $(\bQ, \bU, \bV$ where $\bQ=\bU+\bV)$,
while not generally conserved (owing to quantum transitions such as
$\bU_1+\bV_1\rightleftharpoons
\bU_2+\bV_2=\bQ$), nevertheless, provides a further description 
of such spacetime
``defects''. Finally, the topology of fundamental fermions is said
to be the same as that of the associated vector-triad 
under the internal transformation $\bF$.
\medskip

\noindent
[6] C.\ Nash and S.\ Sen, \emph{Topology and Geometry for Physicists},
Academic Press, New York, 1983.
\medskip

\noindent
[7] T.\ Kajita, for the Super-Kamiokande, Kamiokande collaboration,
hep-ex/9810001.
\medskip

\noindent
[8] M.\ Gronau, \emph{Patterns of Fermion Masses, Mixing Angles and CP
Violation},
in The Fourth Family of Quarks and Leptons, First International Symposium,
edited by: D.B. Cline and Amarjit Soni, Annals of The New York Academy of
Sciences, New York, New York, Volume 518, 1987, p. 190. In the case of
strongly-interacting quarks,
topological constraints of the kind considered here
can, at most, play a minor role in determining
such things as
KM-type matrix elements.  For example, it is well known that the angle
$\theta_c$,
$V_{12}$ and the $d$, $s$ quark masses $m_d$ and $m_s$, respectively,
are related via
$V_{12}=\sin\theta_c=\sqrt{\frac{m_d}{m_s}}$, where $m_s>m_d\ne 0$,
even though the $d$ and $s$ quarks are characterized, like the $\nu_e$
and ($\nu_\mu$  or $\nu_\tau$) neutrinos,
respectively, by distinct 
topologies.  Somehow, the underlying topology 
plays a major role in mixing,
in the case of the weakly-interacting \emph{neutrinos}, but not in the case of
the strongly-interacting 
\emph{quarks}. The small value, and near degeneracy, of neutrino masses is
probably a major factor in explaining such differences between quarks and
neutrinos.
\medskip

\noindent
[9] H.\ Georgi and S.\ L.\ Glashow, ``Neutrinos on Earth and in the
Heavens,'' hep-ph/9808293, page 5, Equation 20.
\medskip

\noindent
[10] If $M(a)=M(\frac{1}{3})$ \emph{any} initial neutrino ``mixture'' at
birth would turn into the mixture $(\frac{1}{3}\nu_e+\frac{1}{3}\nu_\mu +
\frac{1}{3}\nu_\tau)$ when detected at a great distance. If $M(a)=M(1)$
\emph{any} initial neutrino ``mixture'' at birth (e.g.,
$\alpha\nu_e+\beta\nu_\mu+\gamma\nu_\tau; \alpha+\beta+
\gamma
=1$) would turn
into the mixture $[\alpha\nu_e+\frac{(\beta+\gamma)}{2}\nu_\mu +
\frac{(\beta+\gamma)}{2}\nu_\tau]$ when detected at great distance.
\medskip

\noindent
[11] Ordinarily one thinks of a maximization process as describing a
condition of \emph{unstable} equilibrium. However, in the present situation,
two \emph{different} kinds of ``forces'' (topological and quantum) are in
opposition. Taken together, these ``forces'' are constantly attempting to
\emph{maintain} the product $f(a)=a\cdot b$ at its \emph{maximum} value.
That is, the ``system'' described by $f(a)$ should be in a state of
\emph{stable} equilibrium.
\medskip

\noindent
[12] To put it another way, $a^{-1}_0$ is the \emph{arithmetic mean} of
$a^{-1}_{\frac{1}{3}}$ and $a^{-1}_1$.
\medskip

\noindent
[13] In Section 3.0, the matrix elements $a$, $b$ and $c$ were calculated
without an in-depth understanding of the underlying dynamics. The present
description, based on a simple harmonic-oscillator ``potential'' $V(x)$,
represents
a very crude first-attempt to remedy this situation. However, strictly 
speaking, to
reach equilibrium at $a=a_0$, it would also be necessary to include
``damping'' of some sort. This has not been done.

\end{document}